\documentclass{article}

\usepackage[a4paper,
            bindingoffset=0.2in,
            left=0.75in,
            right=0.75in,
            top=1in,
            bottom=1in,
            footskip=.25in]{geometry}

\usepackage[most]{tcolorbox}
\newtcolorbox{myquote}[1][]{%
    colback=black!5,
    colframe=black!5,
    notitle,
    sharp corners,
    borderline west={2pt}{0pt}{red!80!black},
    enhanced,
    breakable,
}

\usepackage{url}
\usepackage[superscript,biblabel]{cite}

\newcounter{daggerfootnote}

\usepackage{pifont}

\newcommand{\xmark}{\ding{55}}

\usepackage{booktabs}
\usepackage{amsmath,amssymb,amsfonts}
\usepackage{graphicx}
\usepackage{array}
\usepackage{float}
\usepackage{caption}
\usepackage{subcaption}
\usepackage{multirow}

\usepackage{wrapfig}
\usepackage{caption}

\usepackage{xcolor}
\usepackage[normalem]{ulem}

\title{CARE: A Multimodal Corpus for Studying Speech\\and Non-Verbal Communication Across Multiple Medical Conditions}
\author{\textit{David Gimeno-Gómez}\textsuperscript{$1$,$\dagger$}, \textit{Catarina Botelho}\textsuperscript{$2$,$3$,$\dagger$},\\\textit{Carlos-D. Martínez-Hinarejos}\textsuperscript{$1$}, \textit{Isabel Trancoso}\textsuperscript{$3$,$4$}, \textit{Alberto Abad}\textsuperscript{$3$,$4$}}
\date{
$^{1}$PRHLT, Universitat Politècnica de València, Spain; $^{2}$Sword Health, Portugal;\\%
$^{3}$INESC-ID, Portugal; $^{4}$Instituto Superior Técnico, Universidade de Lisboa, Portugal
}

\begin{document}
\maketitle

\begingroup
\renewcommand{\thefootnote}{\fnsymbol{footnote}}
\footnotetext[2]{These authors contributed equally to this work.}
\endgroup

\begin{abstract}
\textit{Automatic analysis of multimodal speech has shown strong potential for computationally detecting and monitoring a wide range of neurological, psychiatric, and respiratory conditions. However, progress in this field is limited by existing publicly accessible datasets, which are often small in scale, focused on a single condition or disease, and primarily speech focused. Moreover, if key confounding variables such as education, medication use, comorbidities, or mood state are insufficiently documented, the reliability and interpretability of computational analyses are further compromised. To address these limitations, we introduce \textbf{CARE v1.0}, a curated multimodal English dataset of approximately 144 hours of short video interviews collected from 612 individuals across 12 medical conditions plus a control cohort. For each video, a comprehensive set of clinically relevant multimodal descriptors is provided, alongside structured metadata covering factors such as medication, life impacts, and expressed emotions. The corpus’s breadth and heterogeneity support a wide range of applications, including automatic disease and symptom detection, multimodal modelling of speech and non-verbal behaviour under emotionally charged contexts, and studies of disease trajectories and coping processes.}
\end{abstract}

\section{Background \& Summary}
\label{sec:background}

The automatic analysis of multimodal speech recordings for the detection and monitoring of speech-affecting conditions has emerged as a promising and rapidly expanding research field~\cite{idrisoglu2023applied}. Numerous studies have demonstrated the potential of automatic methods not only in speech and language impairments, such as stuttering~\cite{bayer2023stuttering} or sigmatism~\cite{krecichwost2021sigmatism}, but also across a broad spectrum of health conditions, including neurological disorders (e.g., Parkinson's~\cite{favaro2023interpretable, botelho2024speech,gimeno2025unveiling}, Alzheimer's~\cite{fraser2015linguistic,botelho2024macro,perez2025automated}, Huntington’s~\cite{yoon2006speech_Huntington}, and Amyotrophic Lateral Sclerosis~\cite{gomez2015ALS}); psychiatric conditions (e.g., depression~\cite{cummins2015review}, psychosis~\cite{de2023acoustic}, and bipolar disorder~\cite{karam2014bipolar}); and respiratory diseases (e.g., obstructive sleep apnea~\cite{botelho2019osa,perero2019adversOsa}, asthma~\cite{alam2022predicting}, 
and COVID-19~\cite{schuller2021covid19}).

In response to this growing interest, several public resources have been released, including community-driven data challenges such as the Computational Paralinguistics Challenge (ComParE)~\cite{IS11-schuller2011interspeech, schuller2015interspeech, schuller2017interspeech, schuller2019interspeech}, the Alzheimer’s Dementia Recognition through Spontaneous Speech (ADReSS) series~\cite{ADReSS, luz2021adress_o, luz2024overview}, the Taukadial~\cite{garcia2024connected}, and the PROCESS~\cite{tao2025process} corpora, as well as datasets not adopted in challenges such as NeuroVoz~\cite{mendes2024neurovoz} (Parkinson’s disease) or the Androids Corpus~\cite{tao2023androids} (depression). A notable recent effort is the Bridge2AI Voice dataset~\cite{bensoussan2025b2aiv}, which offers large-scale coverage of multiple health conditions across adults and children.
Also worth mentioning is the Crowdsourced Language Assessment Corpus (CLAC)~\cite{haulcy2021clac}, which comprises recordings of healthy speakers performing standardized tasks designed to elicit speech and language features.
In spite of their relevance, these resources are limited to audio recordings only and their corresponding text transcriptions.

However, human communication is inherently multimodal, encompassing a range of non-verbal behaviors that accompany spoken language. Long-standing research has shown that body language, such as arm and hand movements, forms an integrated system with speech~\cite{mcneill1992hand,kendon2004gesture}. More recently, studies have also highlighted the communicative role of head movements, gaze shifts, and upper-body motion, which also contribute to this coordinated multimodal system~\cite{wagner2014overview,charuau2027adaptive}.
These co-speech gestures not only can convey explicit semantic meaning, but they frequently align with the rhythm and emphasis of speech, serving as expressive channels for emotional states~\cite{noroozi2021emobody}. While well established clinically, co-speech and other non-verbal signals are increasingly leveraged in computational modeling to capture emotional state, cognitive load, and neurological or psychiatric conditions~\cite{turaev2023bodylang}.
For example, recent studies have highlighted the relevance of detecting hypomimia in Parkinson’s disease~\cite{riosurrego2025hypomimia}, reduced affective expressivity and atypical head-movement patterns in psychosis~\cite{martin2024behavioral}, altered blinking and downward head-pose inclination in depression~\cite{fiquer2013talking,gimeno2024reading}, and gaze abnormalities in Alzheimer's~\cite{beltran2018eyemovements}. 

Thus, diverse efforts have extended speech-centered datasets towards multimodal analysis. The recent ParkCeleb dataset~\cite{favaro2024unveiling} includes 40 celebrities who publicly disclosed their Parkinson’s disease diagnosis, providing up to 60 hours of video data alongside a matched control group. Nonetheless, the recordings are highly variable and ``in-the-wild" in nature, including studio interviews, press conferences, red-carpet events, and public speeches, which limits their suitability for controlled multimodal biomarker studies. A well-established and representative benchmark in this area is the Audio/Visual Emotion Challenge (AVEC) series~\cite{AVEC2016,ringeval2019avec,ringeval2018avec}, which has promoted the development of multimodal methods for the automatic detection of mood disorders such as depression and bipolar disorder.

However, despite these remarkable prior efforts, progress in computational approaches to multimodal speech-based health assessment remains hindered by 
limitations in available resources, including small dataset sizes, the focus on single conditions or diseases, and the absence or insufficent documentation of key confounding variables such as education, medication use, comorbidities, or mood state~\cite{botelho2024speech}.

\vspace{0.15cm}
\noindent\textbf{Contributions.} In this work, we introduce \textbf{CARE v1.0} (\textbf{C}onversational \textbf{A}udio-visual \textbf{R}ecordings of health \textbf{E}xperiences), a curated multimodal resource for benchmarking speech and non-verbal behavioural analysis in health-related communication, derived from the Health Experience Insights (HEXI) platform~\cite{ziebland2021polyphonic}. The resulting corpus contains 622 profiles representing 612 unique individuals, classified as either patients or controls. It spans 12 medical conditions with known or expected relevance to speech and non-verbal communication -- asthma, chronic pain, cleft lip and palate, COVID-19, depression, epilepsy, fibromyalgia, lung cancer, motor neuron disease, Parkinson’s disease, psychosis, and stroke -- along with control participants who are often caregivers, bereaved relatives, or healthcare and research professionals discussing their experiences supporting affected individuals. In total, the dataset comprises 4,281 short video clips amounting to 143.5 hours of material. For each video clip, a comprehensive collection of pre-computed multimodal descriptors capturing speech production, facial activity, gaze patterns, and body movements is provided, alongside demographic information and additional structured metadata automatically extracted from narrative accounts, such as ethnic background, medication or treatment references, life impacts, potential comorbidities, and expressed emotions. \textbf{CARE v1.0} therefore provides a rich and versatile resource for advancing research on multimodal digital biomarkers, enabling investigations into automatic disease and symptom detection, as well as other directions such as modelling speech- and behaviour-derived signals under emotionally charged conditions, and computational analyses of disease trajectories and coping processes. We anticipate that the breadth and heterogeneity of this corpus will support a wide range of studies, including many applications beyond those currently envisioned.

\section{Methods}
\label{sec:methods}

\subsection{Input Data}

The data used in this study is obtained from the HEXI platform (\url{https://hexi.ox.ac.uk/}), a publicly accessible archive that, over the past two decades, has collected a selected set of short video excerpts from interviews with patients and caregivers sharing their experiences of illness, treatment, and care, accompanied by thematic analyses. As stated by the platform, its purpose is to enable contributors to:

\begin{center}
“\textit{share their experiences of health on film to help others understand\\what it is really like, from people who have been there.}”
\end{center}

At the time of the corpus construction, HEXI hosted interviews spanning a broad range of health conditions and topics, listed on the platform’s publicly available A–Z index (\url{https://hexi.ox.ac.uk/a-z}). Each specific condition/topic comprises multiple participant profiles, where each profile includes short video interview clips, verbatim transcripts, a written narrative summary, and demographic information. This complete set of HEXI topics served as the initial input dataset from which our corpus was subsequently curated, as described in the following Subsection~\ref{sec:construction}.

\vspace{6pt}
\noindent\textbf{Ethics Statement.} The HEXI platform~\cite{ziebland2021polyphonic} is supported and developed through the efforts of the Medical Sociology \& Health Experiences (MS\&HERG) research group at University of Oxford. The studies are approved by the Multi-centre Research Ethics Committee (MREC) and the Eastern MREC. The archive and associated materials are owned and maintained by the University of Oxford and are made available for research, teaching, and scholarly use.

\subsection{Corpus Construction}
\label{sec:construction}

The construction of the corpus followed a two-stage workflow, consisting of a data curation phase in which health conditions and participants of interest were selected, followed by a metadata retrieval process in which clinically and analytically useful descriptors were automatically extracted.

\vspace{6pt}
\noindent\textbf{Data Curation.} As mentioned above, the corpus was derived from the full set of HEXI topics. Within these topics (e.g., specific conditions or health-related experiences), each participant contributes one or more video clips extracted from interview recordings. The collection of all materials associated with a single participant constitutes a \textit{profile}. Importantly, not all topics were relevant to our research focus, and not all participant profiles contained usable video or textual data suitable for computational and machine-learning analyses. Therefore, our methodology consisted of two main stages:

\begin{itemize}
    \item \textit{Condition Selection.} We began with the original full list of conditions and topics available on  HEXI. Notably, we distinguished between two types of profiles: those corresponding to \textit{patients} and those corresponding to \textit{controls}. For \textit{patient} profiles, we selected conditions based on literature evidence of known or expected relevance to speech and non-verbal communication. For \textit{control} participants, we focused on health-related topics discussed by caregivers, bereaved relatives, or healthcare and research professionals. These topics were selected to span diverse situations and experiences, not necessarily limited to the same conditions selected for patients.

    \item \textit{Profile Filtering.} Within each selected condition, subject profiles were further screened. Specifically, we retained profiles that contained at least one complete video recording, provided that the subject's narrative account was also available. Profiles lacking usable video material -- due to data corruption, privacy restrictions, or technical artifacts -- were excluded, even when audio was present, to preserve the consistency of our multimodal dataset. Additional exclusion criteria included blacked-out videos, AI-generated or synthetic looping footage, and recordings performed by actors.
    
\end{itemize}

After this selection process, our curated corpus included  445 patient profiles spanning 12 different conditions and 177 control profiles. We note that, within several conditions, subgroups were identified reflecting differences in symptoms, treatments, or the specific topics discussed during the interviews. How these subgroups were defined is described in detail in Section~\ref{sec:record}.

\vspace{6pt}
\noindent\textbf{Metadata Retrieval.} To enrich the dataset, we automatically extracted structured metadata from participants’ narrative accounts collected through conversations with social science researchers. These metadata fields provide additional contextual information -- such as medication references, life impacts, or emotional states -- that can support downstream clinical and computational analyses.

Structured metadata were  extracted by prompting \textit{gpt-oss-120b}~\cite{openai2025gptoss120bgptoss20bmodel} with the corresponding narrative descriptions and querying for predefined fields. The queried fields differed depending on whether the participant was categorized as a control or a patient. The selection of \textit{gpt-oss-120b} was informed by a validation study in which we compared the performance of several open-source Large Language Models (LLMs) on a subset of manually annotated samples (see Section~\ref{sec:validation} for details). To ensure deterministic and reproducible outputs, all model interactions were executed using the vLLM~\cite{kwon2023efficient} engine, with the temperature parameter fixed to 0 for all queries.

The LLM outputs were then post-processed and converted into structured metadata tables, released as separate CSV files corresponding to the patient and control cohorts. In both cases, fields allowing multiple annotations were represented as lists of strings, while missing values were imputed with ``na". It is important to note that the model was explicitly instructed to return ``na" when a field was not applicable or when insufficient information was available to provide a reliable response.

\section{Data Record}
\label{sec:record}

This section describes the hierarchical structure of the \textbf{CARE v1.0} corpus, comprising the metadata, annotations, and a comprehensive set of pre-computed multimodal descriptors.

\begin{figure}[ht]
    \centering
    \includegraphics[width=0.8\textwidth]{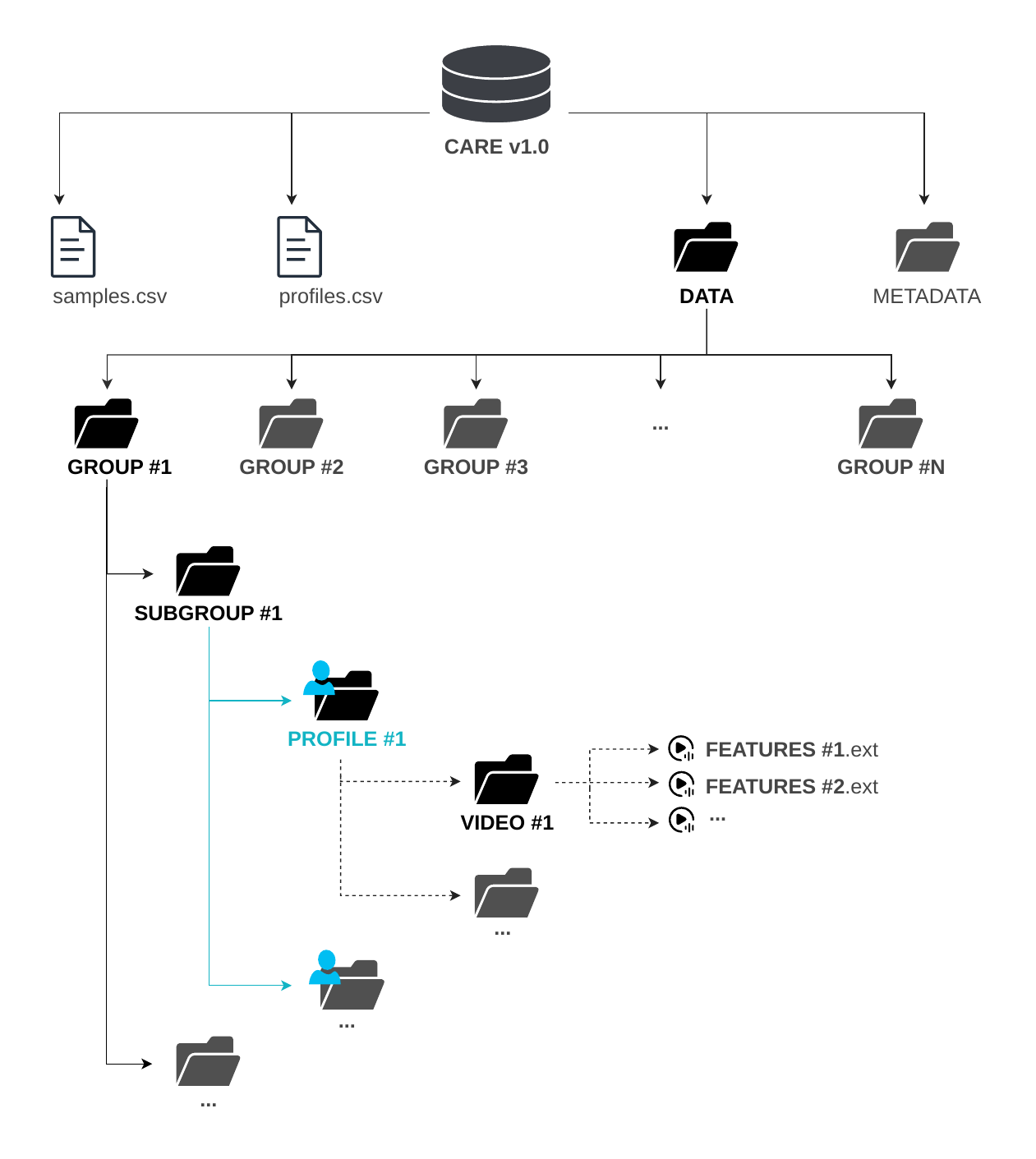}
    \caption{\textbf{Hierarchical representation of the data} records and their organization.}
    \label{fig:record}
\end{figure}

\vspace{6pt}
\noindent\textbf{Overview.} As shown in Figure~\ref{fig:record},
the dataset is organized into two top-level folders and two CSV files. The CSV files include both sample-level information for all 4,281 entries (\texttt{samples.csv}, $\sim$9MB) and profile-level descriptors for all 622 selected profiles (\texttt{profiles.csv}, $\sim$2MB), such as age, sex, and ethnic background. The two top-level folders contain the extracted multimodal features (\texttt{DATA/}, $\sim$119GB)) and two additional CSVs (\texttt{METADATA/}, $\sim$232kB) comprising the structured metadata extracted using LLMs separately for patients and controls. The complete \textbf{CARE v1.0} corpus is distributed through the official Hugging Face Dataset repository~\cite{care2026hf}, following the organization described in this section.

Concerning video-derived descriptors, original video frame rates ranged from 15 to 60 fps (mean 25.6 fps, standard deviation 4.3), with approximately 96\% of the recordings acquired at 25 fps. Similary, although image resolutions varied substantially, with the smallest clip at a resolution of 352$\times$288 pixels and the largest reaching full HD resolution (1920$\times$1080), approximately 68\% of the recordings were acquired at 360$\times$288 pixels. This moderate variability supports research on robust multimodal biomarker modeling under semi-controlled acquisition conditions. No additional normalization of frame rate or spatial resolution was required, as the employed visual feature extraction toolkits operate directly on the original video frames. In contrast, for audio-derived descriptors, all signals were uniformly resampled to 16 kHz and converted to single-channel (mono), as most audio feature extraction toolkits required.

\vspace{6pt}
\noindent\textbf{Data Structure.} Across the dataset, a consistent nomenclature is used, which applies both to the column headings of the metadata and to the overall folder structure:

\begin{itemize}
    \item \underline{\texttt{GROUP}}: denotes the health condition or topic associated with a participant profile, including the control population. \textbf{It does not necessarily correspond to the clinical label} of the speaker. For example, control participants may be assigned to a condition-specific group (e.g., \texttt{PARKINSON}) when describing their experience caring for an affected individual.
    
    \item \underline{\texttt{SUBGROUP}}: specifies a subdivision within a given group. For example, the group \texttt{DEPRESSION} includes subgroups such as \texttt{ANTIDEPRESSANTS} and \texttt{YOUNG-LOWMOOD}, referring to discussions focused on antidepressant treatment or on young adults with low mood, respectively. \textbf{Not all groups contain subgroups}. In such cases, to maintain structural consistency, the subgroup name matches the group name. 

    \item \underline{\texttt{PROFILE}}: is an integer identifier assigned to each participant profile. Remarkably, a participant may have multiple profiles in different groups or subgroups. To support identification of such cases, we provide metadata (the \texttt{duplicated} field) that links profiles corresponding to the same participant. Profile-level granularity is retained to enable analyses of co-morbidity and cases where the same individual appears in multiple contexts -- for example, as a patient in one profile and as a caregiver in another.
\end{itemize}

\sloppy{
Based on these concepts, each video sample is uniquely identified by its position in the directory hierarchy, following the structure \texttt{GROUP/SUBGROUP/PROFILE\_ID/VIDEO\_ID/}. Here, \texttt{VIDEO-ID} denotes a consecutive integer corresponding to the video number within a participant's profile directory rather than a global dataset-wide index, while both \texttt{PROFILE-ID} and \texttt{VIDEO-ID} are represented as zero-padded three-digit integers. For convenience, this hierarchy can also be expressed as the canonical identifier \texttt{GROUP\_\_SUBGROUP\_\_PROFILE\_ID\_\_VIDEO\_ID}. For example, \texttt{DEPRESSION\_\_ANTIDEPRESSANTS\_\_PROFILE\_296\_\_VIDEO\_000} identifies the first interview video associated with participant 296 within the depression condition and antidepressant intake subgroup. Within each video directory, individual feature files (\texttt{FEATURES-ID}) are named according to the modality or descriptor they contain, while the enclosing directory uniquely identifies the recording across all modalities.}

\begin{table}[ht]
\centering
\begin{tabular}{p{3cm}p{11cm}}
\toprule
\textbf{Metadata File} & \textbf{Fields} \\
\midrule
\texttt{samples.csv}  & label\_id, group\_id, subgroup\_id, profile\_id, video\_id, video\_duration, video\_fps, video\_framecount, video\_resolution \\ \midrule
\texttt{profiles.csv}  & label\_id, group\_id, subgroup\_id, profile\_id, age, sex, age\_of\_diagnosis, remarks, duplicated \\
\bottomrule
\end{tabular}
\caption{\textbf{Overview of metadata fields} at sample and profile levels.}
\label{tab:metadata-overview}
\end{table}

\vspace{6pt}
\noindent\textbf{Profile \& Sample Metadata.} As outlined in Table~\ref{tab:metadata-overview}, the metadata associated with the corpus are organised into two complementary tables. The \texttt{samples.csv} file contains one entry per video recording, including identifiers and technical recording characteristics (e.g., duration, frame rate, and resolution). The \texttt{profiles.csv} file, on the other side, contains one entry per participant profile, providing demographic and clinical information such as age, sex, diagnosis-related variables, and profile annotations.

\begin{table}[ht]
\centering
\begin{tabular}{p{3cm} c c p{7.5cm}}
\toprule
\textbf{Metadata Type} & \textbf{Patients} & \textbf{Controls} & \textbf{Fields} \\
\midrule
Demographics  & \checkmark & \checkmark & Name, Ethnic background, Sex, Job/Occupation \\
Clinical & \checkmark & \xmark & Main medical condition, Time since condition onset, Current symptoms, Initial symptoms, Current medication, All medications or treatments mentioned, Other strategies for improving, Comorbidities, Assistive devices (voice software or ventilator) \\
Psychosocial & \checkmark & \xmark & Mood mental health or other consequences \\
Caregiving & \xmark & \checkmark & Role, Relationship to care recipient, Diagnosis of the care recipient, Years of care, Consequences for personal life \\
Emotional & \checkmark & \checkmark & Main emotions (Ekman’s eight categories~\cite{ekman1992argument}) \\
\bottomrule
\end{tabular}
\caption{\textbf{Overview of structured metadata fields} for patient and control profiles.}
\label{tab:structured-overview}
\end{table}

\vspace{6pt}
\noindent\textbf{Structured Metadata.} As previously introduced, each participant's profile included an extensive narrative providing a chronological account of experiences and relevant life events, as documented during the interviews with the social scientist researcher. Depending on the participant group, these narratives reflect either the perspective of the patient experiencing the condition or, in the case of control participants, that of a caregiver describing the person they support as well as their own caregiving experience. These narratives served as the primary text source for our automatic LLM-based structured metadata retrieval using \textit{gpt-oss-120b}~\cite{openai2025gptoss120bgptoss20bmodel}.

As summarized in Table~\ref{tab:structured-overview}, the retrieved fields differ between patient and control profiles, with 15 fields defined for patients and 10 for controls, of which five are shared. Apart from demographics, only the emotional metadata is common to both profiles, which is notably constrained to eight categories based on Ekman’s widely used framework for basic emotions~\cite{ekman1992argument} that is commonly adopted in emotion-recognition research: neutral, happiness, sadness, surprise, fear, disgust, anger, and contempt. Besides, we also predefine the possible roles associated with \text{control} profiles, namely: \textit{caregivers and supporters}, which include those providing direct or emotional support to someone with a disease, for example a family member, a friend or caregiver; \textit{bereaved members}, which include individuals who have lost someone to the disease; \textit{healthcare and research professionals}, who may describe clinical or scientific aspects of the condition; and \textit{advocates}, who promote awareness, support, or representation of people affected by a disease. For all metadata fields, the value ``na" is assigned when the model determines that the information is insufficient or not applicable for a given profile.

\vspace{6pt}
\noindent\textbf{Multimodal Descriptors.} Each participant profile included one or more video excerpts derived from interviews conducted by social scientist researchers. Individual clips are typically centred on a specific topic relevant to the participant's experience. In most cases, the participant's face and upper body remain continuously visible throughout the recording, although brief periods in which the participant is outside the camera view may occasionally occur. In a small number of cases, the corpus also includes self-recorded video segments contributed by participants, which may exhibit different camera viewpoints.

Regarding the audio modality, the participant is consistently the predominant speaker, although brief interventions from the social science researcher (interviewer) are commonly present.

Given these characteristics, the extraction of multimodal descriptors was designed to account for the conversational nature of the recordings. To minimise the influence of occasional interviewer interventions, speaker diarization outputs generated with DiariZen~\cite{han2025diarizen} were first used throughout the audio processing pipeline to identify participant speech segments prior to feature extraction. These diarization outputs are also included in the released dataset, enabling users to investigate conversational phenomena, such as turn-taking patterns, where present. Based on this processing pipeline, we computed a comprehensive set of selected speech and non-verbal multimodal descriptors aligned with the participant's conversational turns to capture complementary behavioural signals across modalities. Broadly, these descriptors can be grouped into four categories:

\begin{itemize}
    \item \textbf{Speech \& Voice.} Audio representations were extracted at multiple levels of abstraction. We provide the eGeMAPS v2.0 feature set~\cite{eyben2016egemaps} on a turn-by-turn basis, including both frame-level low-level descriptor (LLD) trajectories and turn-level functional summaries, capturing spectral and voice quality characteristics commonly used in affective and clinical speech analysis. Additional articulation, prosody, and phonation descriptors extracted with DisVoice~\cite{vasquez2020disvoice} encompass a broader range of measures related to speech production, vocal quality, and pausing behaviour. Deep speech embeddings are also provided using the multilingual XLS-R wav2vec 300M~\cite{babu2202wav2vec} and TRILLsson 4.1~\cite{shor2022trillsson} models, aggregated at participant turn level. Following the same segmentation strategy, emotion-related trajectories were extracted to characterise the temporal evolution of arousal, valence, and dominance throughout each recording, together with their corresponding latent embedding representation~\cite{wagner2023dawn}. For these deep learning representations, only aggregated embeddings are released to reduce the risk of recovering sensitive linguistic information or reconstructing the underlying speech signal.
    
    \item \textbf{Facial Expression \& Gaze.} Most facial representations were extracted using OpenFace v2.2.0~\cite{tadas2018openface}. The resulting features include 68 facial landmark coordinates capturing facial geometry, together with intensity values of key facial action units used to characterise clinically-aligned expressive behaviour. In addition, 28 eye landmarks and gaze direction vectors were computed to capture visual attention and eye movement patterns. Deep facial embeddings were also extracted using EmoNet~\cite{toisoul2021emonet} to provide deep latent representations of affective facial expressions. Furthermore, EmoNet-derived emotion trajectories were estimated frame-wise, including continuous valence and arousal dimensions as well as the probabilities associated with the eight Ekman emotion categories, thus enabling the analysis of the temporal evolution of facial affect throughout each recording. Frames in which no face was detected are explicitly annotated to enable robust handling of missing visual information during downstream analysis. Finally, blink events were estimated using an Eye Aspect Ratio (EAR)-based approach~\cite{soukupova2016real} computed from eye landmarks, providing an additional indicator of ocular activity. With the exception of the EmoNet latent embeddings, which are aggregated at the participant-turn level, all visual descriptors are provided frame-wise and can subsequently be segmented according to the released speaker diarization boundaries.
 
    \item \textbf{Head \& Body Movement.} Head pose and rotation estimates were also extracted using OpenFace v2.2.0, providing frame-wise yaw, pitch, and roll angles relative to the camera coordinate system. In addition, upper-body movement was estimated using MediaPipe~\cite{hongyi2020ghum}, providing 33 body keypoints that describe posture and gross movement patterns over time. Together, these representations characterise psychomotor manifestations, namely head orientation and upper-body motion dynamics, during the interviews and enable the derivation of higher-level features such as motion intensity, postural variability, and coordination between head and body movements for downstream analyses.

    \item \textbf{Linguistics.} The manually curated transcripts provided by HEXI included rich transcript annotations identifying non-verbal vocal events and indicators of disfluency and interactional behaviour.  Leveraging this information, we computed a set of transcript-derived linguistic descriptors, including total word count, speech rate, lexical diversity computed over content words (excluding stopwords), content-word repetition ratio, filler and backchannel frequencies, as well as counts of the annotated non-verbal vocal events (e.g., laughter, crying, and breathing). Importantly, the transcripts were temporally aligned to the corresponding audio using WhisperX~\cite{bain2023whisperx}, enabling the computation of these descriptors at both the video and participant-turn levels. Together, these features provide lightweight yet informative characterisations of verbal production that complement the acoustic and visual modalities. In addition to these handcrafted descriptors, each participant turn is represented by a dense semantic embedding extracted using the \texttt{all-mpnet-base-v2} Sentence Transformer model~\cite{song2020mpnet}. These embeddings provide a compact neural representation of the turn semantics, analogous to the latent representations extracted from the acoustic and visual modalities.

\end{itemize}

This comprehensive set of multimodal descriptors enables the study of speech, facial, and body language behaviour from the perspective of verbal and non-verbal communication in digital health, supporting a wide range of downstream computational analyses. Full documentation of all features, including detailed definitions, dimensionalities, and summary statistics, is provided in the accompanying repository.

\section{Data Overview}
\label{sec:overview}

Our dataset encompasses video samples collected across a diverse set of medical and mental-health conditions, selected for their potential to manifest measurable changes in speech and body language. Conditions were organized into four categories: \textbf{(i) neurological or motor disorders} (\textit{Parkinson’s disease}, \textit{motor neuron disease}, \textit{stroke}, \textit{epilepsy}), involving impaired motor control that can disrupt articulation, fluency, and coordinated facial or gestural expression; \textbf{(ii) respiratory or structural conditions} (\textit{asthma}, \textit{COVID-19}, \textit{lung cancer}, \textit{cleft lip and palate}), primarily affecting voice quality, breathing patterns, and articulatory precision; \textbf{(iii) chronic pain or fatigue-related conditions} (\textit{chronic pain}, \textit{fibromyalgia}), associated with reduced energy, slower speech rate, and subtle changes in facial expressiveness or body posture; and \textbf{(iv) mental-health disorders} (\textit{depression}, \textit{psychosis}), involving altered prosody, emotional expressiveness, coherence, and head-tilt or other non-verbal cues.

\begin{figure}[ht]
    \centering
    \includegraphics[width=0.95\textwidth]{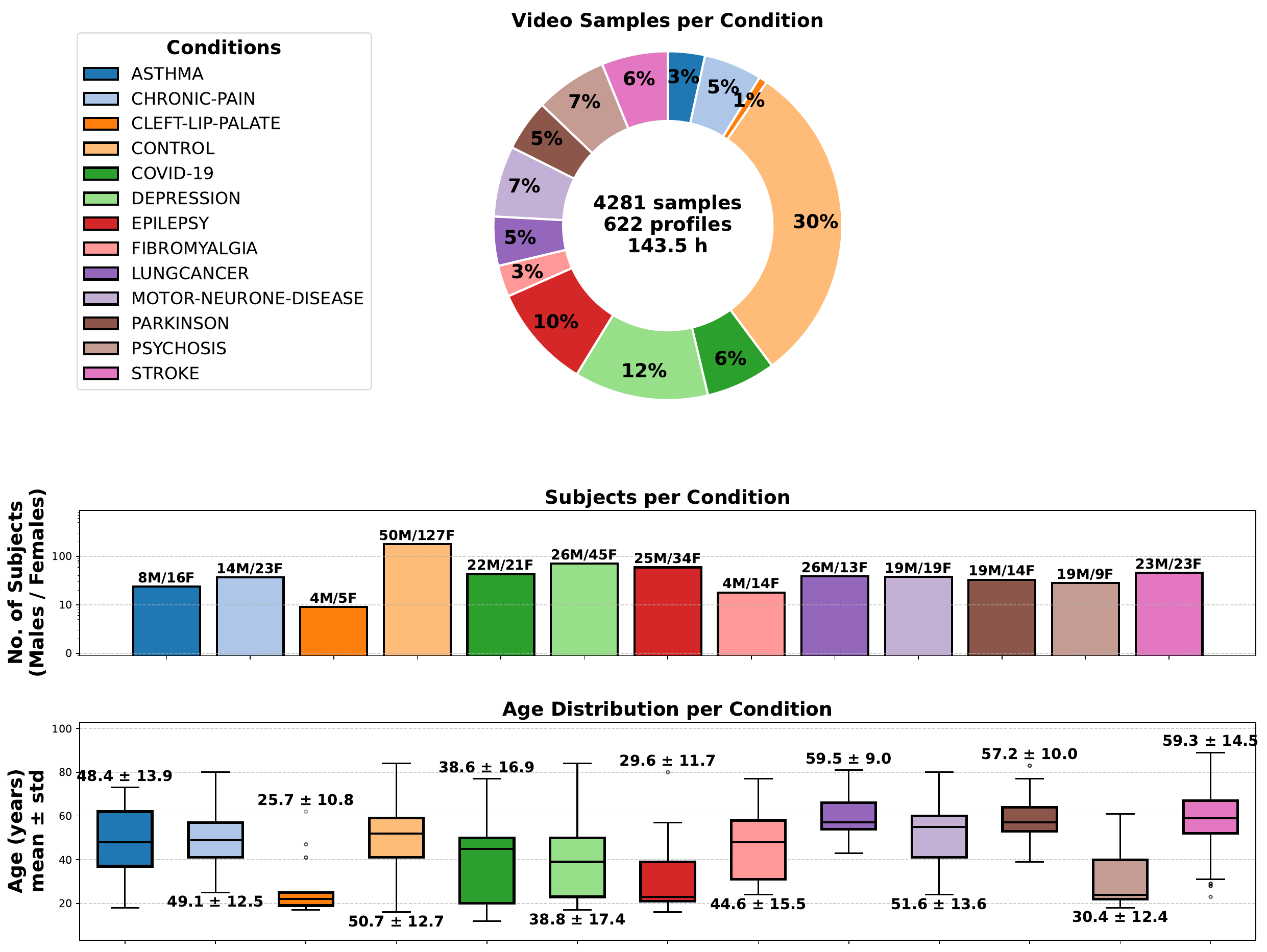}
    \caption{\textbf{Overview of the dataset across medical conditions.} Top: distribution of video samples per condition (percentages shown; center indicates total number of samples, unique subjects, and total recording duration). Middle: number of unique subjects per condition (logarithmic scale), with male (M) / female (F) counts indicated. Bottom: age distribution per condition, with total mean ± standard deviation.}
    \label{fig:overview}
\end{figure}

Together with the control group, the dataset comprises 4,281 video samples, with an average of 10.9 {\small $\pm$9.3} participant turns per video. As reflected in Figure~\ref{fig:overview}, these samples are distributed unevenly across the 13 distinct conditions. Control participants account for the largest share (30\%), while some conditions, such as cleft lip and palate, are minimally represented (1\%). This uneven distribution may limit certain analyses, particularly for underrepresented conditions, reflecting a common challenge in real-world applications where data is often scarce. Nevertheless, the dataset’s coverage across multiple conditions and the inclusion of participants with potential comorbidities offer opportunities to investigate the study of cross-disease patterns and multimodal cues that may help mitigate the limitations posed by such data-scarce clinical contexts.

Across the 622 participant profiles, the mean proportion of female participants across conditions is 54.9\% {\small $\pm$13.2}, indicating a slightly higher representation overall, but with some substantial variability across conditions. Notably, the control and fibromyalgia groups have a female proportion exceeding 70\% on average, whereas psychosis and lung cancer groups include only about 30\% females. Regarding age, participants span a wide range, with a mean of 48.1 {\small $\pm$17.3} years, covering much of the adult lifespan. Among the youngest participant populations are those with cleft lip and palate, epilepsy, and psychosis. Overall, these characteristics highlight the dataset’s heterogeneity, supporting analyses across diverse conditions, age groups, and sexes.

\begin{figure}[ht]
    \centering
    \includegraphics[width=0.95\textwidth]{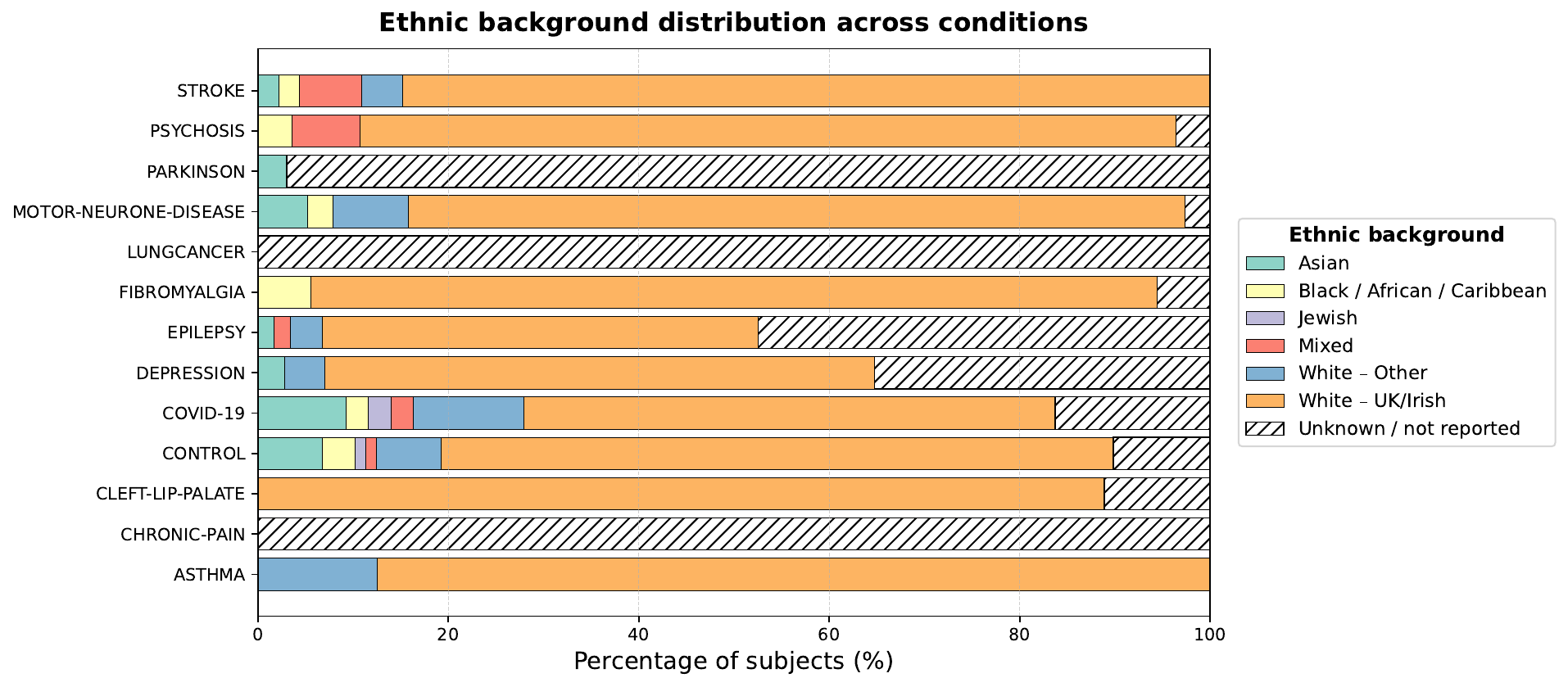}
    \caption{\textbf{Ethnic and nationality backgroun metadata distribution across conditions.} The dataset reflects some diversity but is skewed toward White UK or Irish participants.}
    \label{fig:ethnic}
\end{figure}

\noindent
\begin{minipage}[t]{0.60\textwidth}
\setlength{\parindent}{1.5em}
Another important aspect of this corpus is that the control group comprises individuals with heterogeneous roles in relation to health and disease. As illustrated in Figure~\ref{fig:controlrole}, most control participants are caregivers and supporters, reflecting sustained engagement through close personal relationships. The inclusion of bereaved members, advocates, and healthcare professionals further extends the capture of experiential perspectives shaped by illness. Therefore, unlike previous corpora, where control participants are often drawn from unrelated domains, our control group contributes data grounded in health-related experiences, providing a more contextually relevant reference population.

Figure~\ref{fig:ethnic} complements these analyses by showing the distribution of ethnic and nationality backgrounds across all conditions. While LLM-based extraction of these attributes was often more precise, we aggregated participants into broad categories for clarity. Overall, the corpus reflects some diversity. However, the majority of participants are White UK or Irish, introducing a skew toward these populations. Notably, most participants are English speakers, though English may not be the first language for some, potentially affecting speech patterns and prosody. Only profile 606, from the stroke group, speaks another language (Punjabi), but all corresponding metadata, including transcripts, were translated into English by HEXI. Other groups, such as Parkinson's and chronic pain, miss this information partially or entirely.
\end{minipage}
\hfill
\begin{minipage}[t]{0.38\textwidth}
\vspace{0pt}
\centering
\includegraphics[width=\linewidth]{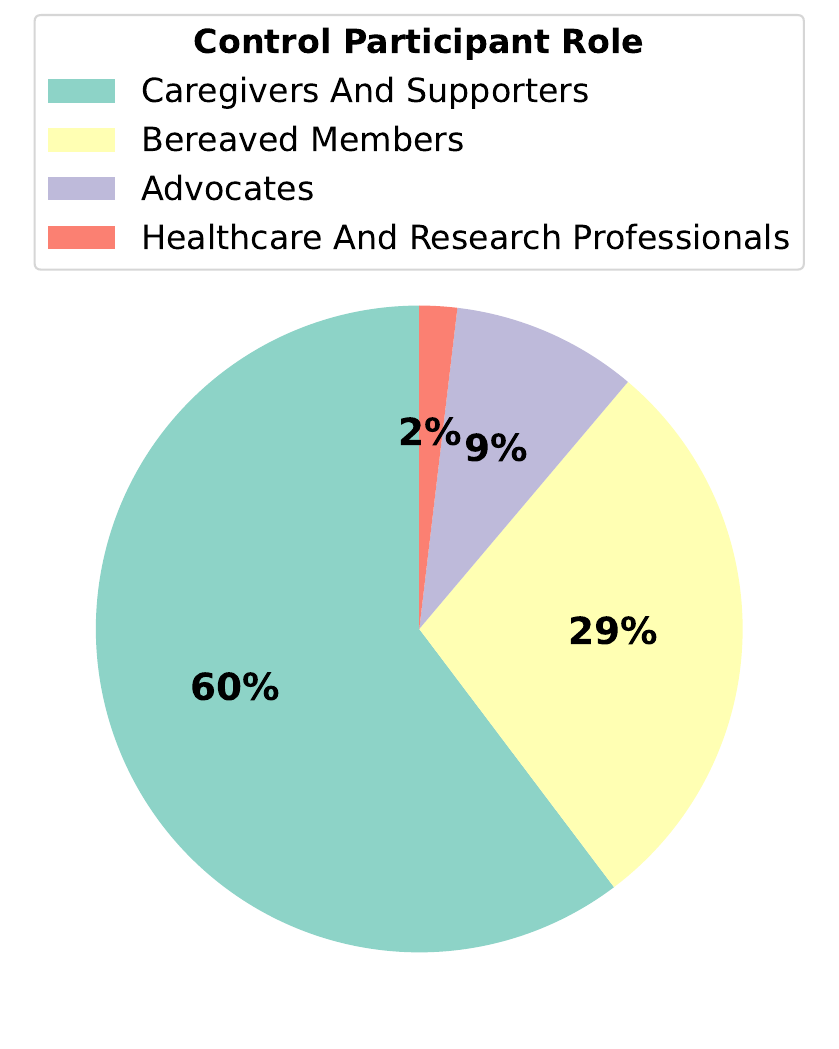}
\vspace{-1cm}
\captionof{figure}{\textbf{Distribution of roles among control profiles}, highlighting the heterogeneous nature of the participants composing this group.}
\label{fig:controlrole}
\end{minipage}

\begin{figure}[ht]
    \centering
    \includegraphics[width=0.8\textwidth]{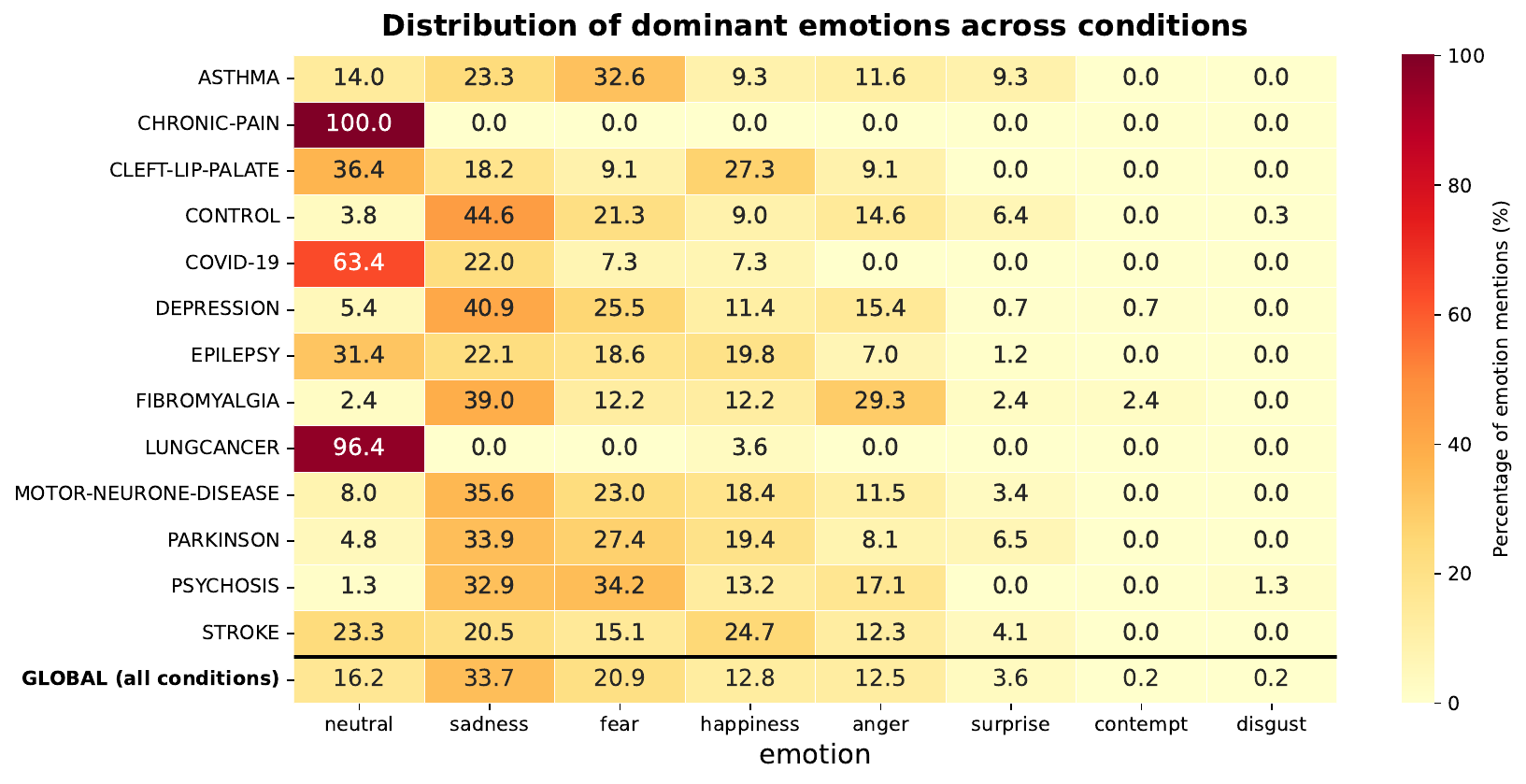}
    \caption{\textbf{Distribution of dominant emotions across conditions derived from textual narratives.} Neutral, sadness, and fear expressions are mostly prevalent across conditions.}
    \label{fig:emotions}
\end{figure}

Among the multi-value structured metadata, we also retrieved dominant emotions expressed in the narratives accounts of participant stories. To analyze this information, we computed emotion frequencies at the participant level and normalized them as percentages per condition, as shown in Figure~\ref{fig:emotions}. A first notable pattern is the prominence of \textit{neutral} or low-emotion annotations across several conditions. This likely reflects the third-person reporting style adopted by researchers and the clinically oriented nature of the narratives, which may attenuate explicit emotional expression. Nonetheless, \textit{sadness} and \textit{fear} emerge as the most prevalent emotions across most conditions, plausibly reflecting the emotional burden associated with illness experiences or vulnerabilities related to economic and institutional support. Importantly, emotion inference in this analysis is based solely on textual narratives, underscoring the value of the dataset’s multimodal nature and motivating future analyses that relate complementary cues from facial expression and voice.

\begin{figure}[H]
    \centering
    \includegraphics[width=0.7\textwidth]{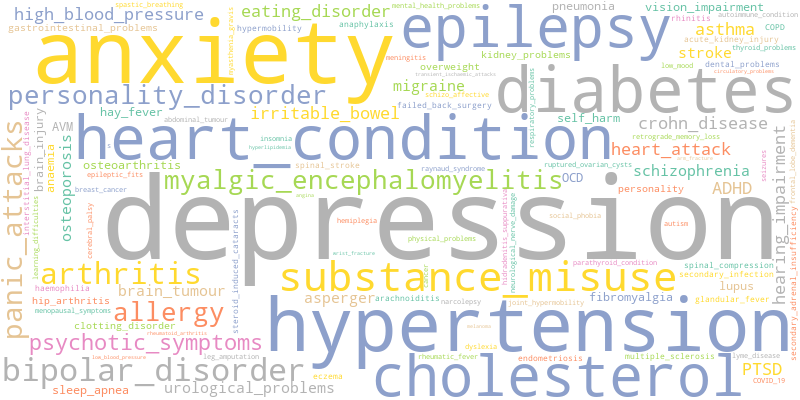}
    \caption{\textbf{Comorbidities distribution.} Word cloud illustrating the main comorbidities reported across patient profiles. Term size reflects relative frequency, highlighting the most prevalent conditions, including depression, anxiety, and diabetes. Multi-word conditions were preserved, and some highly detailed terms were normalized into broader umbrella conditions for clarity.}
    \label{fig:comorbidities}
\end{figure}

From the same narratives, but focusing exclusively on the patient group, a wide range of comorbidities was identified across participants. As reflected in Figure~\ref{fig:comorbidities},  conditions such as anxiety, hypertension, epilepsy, diabetes, depression, and heart-related issues were most frequently observed. This information is crucial as it provides context for interpreting behavioral and emotional patterns, enabling analyses that account for the potential influence of co-occurring medical or mental health disorders both across and within conditions.

\section{Technical Validation}
\label{sec:validation}

To assess the reliability of the automatically generated annotations, we compared the performance of five well-established LLMs using a manually annotated subset of the dataset.

\vspace{6pt}
\noindent \textbf{Manually annotated data subset.} This subset contained 34 profiles: 10 \textit{control} profiles (one from each of the 10 \textit{group-ids} that contain \textit{control} profiles) and 24 \textit{patient} profiles (two profiles for each non-control \textit{label-id}). This sampling strategy ensured diversity across conditions and profile types. The subset contained 14 male and 20 female profiles. 
The automatically annotated dimensions correspond to the metadata fields described in Section~\ref{sec:record}, with the addition of the subject’s name to encourage the model to identify the main subject before completing the requested annotations. Importantly, because several of these metadata fields -- particularly those related to emotions -- require interpretation, the manual annotations (done by a single non-clinic annotator) should be regarded as a reference point rather than an absolute ground truth.

\vspace{6pt}
\noindent \textbf{Models.}
We evaluated five LLMs: \textit{Llama-3.1-8B}~\cite{touvron2023llama}, \textit{Llama-3.3-70B}~\cite{touvron2023llama}, \textit{Qwen-2.5-70B}~\cite{qwen2,qwen2.5}, \textit{Qwen-3-Next-80B}~\cite{qwen2.5-1m,qwen3technicalreport}, and \textit{gpt-oss-120b}~\cite{openai2025gptoss120bgptoss20bmodel}. Experiments were also conducted with \textit{Llama-3.2-3B}~\cite{touvron2023llama}, but its substantially lower performance led us to omit it from detailed reported results.

During preliminary experiments, models from the Llama and Qwen families frequently failed to restrict emotion predictions to the predefined set of Ekman's eight categories~\cite{ekman1992argument}. Attempts to enforce valid outputs using guided decoding with a JSON schema increased hallucinations and reduced overall performance. For conciseness, schema-guided results are not reported.

\begin{table}[ht]
    \centering
    \resizebox{\linewidth}{!}{
    \begin{tabular}{ll|ccc|ccc|ccc|ccc|ccc}
    \toprule
 &  & \multicolumn{3}{c|}{\textbf{Llama-3.1-8B}} & \multicolumn{3}{c|}{\textbf{Llama-3.3-70B}} & \multicolumn{3}{c|}{\textbf{Qwen-2.5-70B}} & \multicolumn{3}{c|}{\textbf{ Qwen-3-Next-80B}} & \multicolumn{3}{c}{\textbf{ gpt-oss-120b}} \\
 &  & ACC & JS & BertS & ACC & JS & BertS & ACC & JS & BertS & ACC & JS & BertS & ACC & JS & BertS \\
 \midrule
\multicolumn{3}{l}{\textbf{Annotations for Controls}} \\
\midrule				
Single- & Name & 0.80 &  &  & 1.00 &  &  & 0.90 &  &  & 1.00 &  &  & 1.00 &  & \\
value & Ethnic background & 1.00 &  &  & 1.00 &  &  & 1.00 &  &  & 1.00 &  &  & 1.00 &  & \\
 & Sex & 1.00 &  &  & 1.00 &  &  & 0.90 &  &  & 1.00 &  &  & 0.90 &  & \\
 & Job or professional occupation & 0.30 &  & 0.81 & 0.40 &  & 0.83 & 0.30 &  & 0.81 & 0.30 &  & 0.82 & 0.30 &  & 0.82 \\
 & Years of care & 0.30 &  &  & 0.70 &  &  & 0.60 &  &  & 0.50 &  &  & 0.70 &  &  \\ 
Multi- & Role & 0.80 & 0.90 &  & 0.80 & 0.90 &  & 0.90 & 0.95 &  & 0.80 & 0.90 &  & 0.80 & 0.90 & \\
value & Relationship to care recipient & 0.90 & 0.90 &  & 0.80 & 0.80 &  & 0.80 & 0.80 &  & 0.90 & 0.90 &  & 0.80 & 0.80 & \\
 & Diagnosis of care recipient & 0.50 & 0.55 & 0.71 & 0.40 & 0.45 & 0.70 & 0.60 & 0.65 & 0.84 & 0.50 & 0.55 & 0.71 & 0.60 & 0.60 & 0.84 \\
 & Consequences for personal life & 0.30 & 0.36 & 0.70 & 0.10 & 0.18 & 0.60 & 0.30 & 0.35 & 0.76 & 0.30 & 0.33 & 0.74 & 0.30 & 0.34 & 0.67 \\
 & Dominant emotions & 0.10 & 0.32 & 0.64 & 0.10 & 0.29 & 0.60 & 0.20 & 0.40 & 0.65 & 0.20 & 0.42 & 0.66 & 0.40* & 0.47 & 0.69 \\

\midrule
\multicolumn{3}{l}{\textbf{Annotations for Patients}} \\
\midrule
Single- & Name & 1.00 &  &  & 1.00 &  &  & 1.00 &  &  & 1.00 &  &  & 1.00 &  & \\
value & Ethnic background & 1.00 &  &  & 1.00 &  &  & 1.00 &  &  & 0.96 &  &  & 1.00 &  & \\
 & Sex & 1.00 &  &  & 0.92 &  &  & 0.96 &  &  & 0.92 &  &  & 1.00 &  & \\
 & Main medical condition & 0.88 &  & 0.94 & 0.92 &  & 0.95 & 0.92 &  & 0.95 & 0.92 &  & 0.95 & 0.92 &  & 0.95 \\
 & Years since condition onset & 0.50 &  &  & 0.67 &  &  & 0.79 &  &  & 0.58 &  &  & 0.88 &  &  \\
 & Voice software or ventilator & 1.00 &  &  & 1.00 &  &  & 1.00 &  &  & 1.00 &  &  & 1.00 &  &  \\
Multi- & Current symptoms & 0.21 & 0.23 & 0.63 & 0.25 & 0.31 & 0.68 & 0.13 & 0.17 & 0.61 & 0.25 & 0.33 & 0.67 & 0.29 & 0.32 & 0.69 \\
value & Initial symptoms & 0.25 & 0.28 & 0.72 & 0.29 & 0.33 & 0.76 & 0.25 & 0.27 & 0.73 & 0.29 & 0.32 & 0.75 & 0.38 & 0.42 & 0.81 \\
 & Current medication & 0.50 & 0.52 & 0.78 & 0.58 & 0.60 & 0.82 & 0.46 & 0.48 & 0.76 & 0.54 & 0.56 & 0.79 & 0.50 & 0.52 & 0.79 \\
 & All medication/treatments & 0.00 & 0.15 & 0.59 & 0.00 & 0.14 & 0.56 & 0.00 & 0.15 & 0.61 & 0.00 & 0.14 & 0.59 & 0.00 & 0.13 & 0.59 \\
 & Other strategies for improving & 0.25 & 0.35 & 0.74 & 0.25 & 0.33 & 0.68 & 0.21 & 0.32 & 0.68 & 0.25 & 0.33 & 0.70 & 0.21 & 0.28 & 0.68 \\
 & Comorbidities & 0.63 & 0.66 & 0.83 & 0.63 & 0.69 & 0.84 & 0.42 & 0.47 & 0.72 & 0.67 & 0.69 & 0.81 & 0.63 & 0.69 & 0.82 \\
 & Job or professional occupation & 0.46 & 0.46 & 0.78 & 0.46 & 0.48 & 0.77 & 0.54 & 0.54 & 0.85 & 0.58 & 0.58 & 0.83 & 0.54 & 0.54 & 0.80 \\
 & Mood, mental health, etc. & 0.21 & 0.24 & 0.57 & 0.17 & 0.19 & 0.57 & 0.21 & 0.22 & 0.62 & 0.17 & 0.21 & 0.57 & 0.25 & 0.27 & 0.61 \\
 & Dominant emotions & 0.04 & 0.25 & 0.64 & 0.08 & 0.21 & 0.63 & 0.25 & 0.47 & 0.71 & 0.33 & 0.53 & 0.76 & 0.29* & 0.46 & 0.70 \\
 \midrule
\multicolumn{2}{r|}{\textbf{Average}}  & 0.56 & 0.44 & 0.72 & 0.58 & 0.42 & 0.71 & 0.59 & 0.45 & 0.73 & 0.60 & \bf 0.48 & 0.74 & \bf 0.63 & \bf 0.48 & \bf 0.75 \\
\bottomrule
\end{tabular}
}
    \caption{\textbf{Performance of the evaluated LLMs on the task of automatic metadata retrieval}, using a manually annotated subset of the dataset as reference. Metrics include accuracy (ACC), Jaccard similarity (JS), and F1 of the Bert Score (BertS). Average scores across all categories are provided in the bottom row. (*) Indicates
    whether emotion predictions were restricted to the predefined set of allowed categories.}
    \label{tab:llm-results}
\end{table}

\vspace{6pt}
\noindent \textbf{Evaluation.} To evaluate metadata extraction, we distinguish between \textit{single-value} fields and \textit{multi-value} fields, the latter represented as lists. 
Exact-match accuracy (ACC) is reported for all fields; however, this metric is overly strict. For multi-value fields it fails to reward partial correctness, and for non-categorical fields it penalizes benign paraphrases.
To mitigate the first limitation, we also compute Jaccard Similarity (JS) used to measure the similarity between two sets, thus capturing partial overlap between predicted and reference sets. To address the second, we compute BERTScore~\cite{zhang2019bertscore} using the model \textit{microsoft/deberta-xlarge-mnli} that showed the highest correlation with human evaluation. We report the F1 form of BERTScore, which penalizes both omissions of reference information and the inclusion of extraneous content.

\vspace{6pt}
\noindent \textbf{Results.} 
Table \ref{tab:llm-results} summarizes model performance across metadata fields defined separately for control and patient participants. Among the evaluated models, \textit{gpt-oss-120b} achieved the highest average performance across ACC, JS, and BERTScore, and was the only model that consistently respected the constraint of selecting emotions exclusively from the seven allowed categories without schema enforcement.
For this reason, the final structured metadata that accompanies this dataset was extracted using \textit{gpt-oss-120b}.

A closer inspection of performance across individual metadata fields reveals that those capturing basic demographic or well-defined clinical attributes, including \textit{name}, \textit{ethnic background}, \textit{sex}, \textit{main medical condition}, and \textit{use of voice software or a ventilator}, exhibited accuracies equal or above 90\%, reaching 100\% for most of these categories. Performance was also high for fields describing the role of the caregiver and the diagnosis of the care recipient, with accuracy around 80\% and JS reaching 90\% for caregiver role.

A different consideration applies to the fields \textit{years of care} and \textit{years since condition onset}, which were ideally numeric. However, in the manually annotated subset, these values were often not provided due to insufficient information, making numeric error metrics (e.g., mean absolute error) unreliable. Consequently, we report accuracy as a more appropriate measure of model performance for these fields. When restricting evaluation to rows where both the reference and the \textit{gpt-oss-120b} output contained numeric values, the mean absolute error was equal or below one year for both fields

In contrast, fields involving more descriptive or contextual information, such as professional occupation, symptoms, or psychosocial consequences, show lower accuracy, largely due to paraphrasing and variable level of detail between manual and model annotations. In these cases, BERTScore provides a more informative measure than accuracy, which was between 0.59 and 0.95. One example is the dominant emotion retrieval, with an overall 0.64 BERTScore in both control and patient groups. For illustrative purposes, Table~\ref{tab:bert-score-examples} presents selected examples of annotations generated by \textit{gpt-oss-120b} alongside the corresponding reference annotations and their BERTScores. Overall, \textit{gpt-oss-120b} achieves a mean BERTScore of 0.75 across all categories. At this level of semantic similarity, the generated annotations closely align with the reference annotations, suggesting adequate performance in capturing the relevant information.

\vspace{6pt}
\noindent\textbf{Full-dataset validation.} While most metadata predictions remained broadly consistent across LLMs, emotion labels were of particular concern due to the constraint that predictions must adhere to a predefined set of allowed categories. For this reason, our validation efforts focused specifically on this dimension, even though \textit{gpt-oss-120b} consistently respected the Ekman categories in the manually annotated evaluation subset. Examination across the full dataset revealed occasional LLM hallucinations. Specifically, a small number of emotion annotations fell outside the Ekman emotion set, including \textit{anxiety} (one profile from the psychosis group), \textit{frustration} (one profile Parkinson’s disease), and \textit{shock} (one control profile), together accounting for less than 1\% of all profiles. In addition, emotion information was missing (``na") for approximately 5\% of the dataset (30 profiles), primarily affecting the chronic pain (3), control (3), COVID-19 (5), epilepsy (5), and lung cancer (11) groups. Despite these minor issues, our experimental results highlight the \textit{gpt-oss-120b} model’s overall reliability in retrieving structured metadata from clinical domains.

\begin{table}[t]
    \centering
    \resizebox{\linewidth}{!}{
    \begin{tabular}{l l l c}
    \toprule
 \textbf{Field} & \textbf{Reference} & \textbf{Model output} & \textbf{BertS} \\
 \midrule
 Job of professional occupation & Nursery nurse & Nurse & 0.7403 \\
 Mood or mental health consequences & Bullying, gained self-confidence & Was bullyed in school but was able to overcome it & 0.5386 \\
 Current symptoms & Occasional stuttering, forgetfulness, fatigue, hesitancy & Occasionally stutters, more forgetful, tired and hesitant & 0.7822 \\
 Other strategies for improving & Breathing exercises, daily walking & Breathing exercises, walks & 0.8539 \\
\bottomrule
\end{tabular}
}
    \caption{\textbf{Examples of semantic matching.} Manual annotations and \textit{gpt-oss-120b} outputs with the corresponding F1 BERT Scores (BertS) across different metadata fields.}
    \label{tab:bert-score-examples}
\end{table}

\section{Usage Notes}

Access to the \textbf{CARE v1.0}~\cite{care2026hf} is available exclusively for non-commercial research use. Researchers wishing to obtain the data must agree to a data use agreement that specifies the terms of use, including, among other conditions, a prohibition on any attempt to re-identify participants, whether directly or through linkage with external data sources. Users must also ensure that any outputs derived from the dataset appropriately cite this manuscript and acknowledge HEXI and the Medical Sociology \& Health Experiences Research Group (MS\&HERG) at the University of Oxford~\cite{ziebland2021polyphonic}.

To support the reproducibility of studies carried out on the \textbf{CARE v1.0} corpus and to encourage the appropriate reuse of the data across diverse research contexts, the following material is provided:

\begin{itemize}
    \item \textbf{Feature Extraction Scripts.} The full set of scripts used to extract all multimodal descriptors from the raw data. These scripts enable users to reproduce the exact feature computation pipeline or to apply it to external datasets, ensuring consistent and fair comparisons across studies.

    \item \textbf{Example Evaluation Workflow.} An example use-case workflow illustrating a participant-level 5-fold cross-validation protocol for a representative binary classification setting (\texttt{CONTROL} versus \texttt{PARKINSON}). The provided example generates profile-level folds while preventing duplicate participant leakage, accounts for key demographic factors such as sex distribution, aggregates turn-level descriptors into profile-level representations, and trains a linear support vector machine (SVM)~\cite{steinwart2008support}. This baseline is intended solely to demonstrate the recommended evaluation workflow, while encouraging users to adopt task-specific models and evaluation protocols appropriate for their applications.
\end{itemize}

Overall, these usage notes are intended to support transparent, reproducible, and consistent use of the CARE v1.0 corpus across a range of downstream analytical settings.

\section{Data Availability}

The \textbf{CARE v1.0} corpus is available for non-commercial research use through the official Hugging Face Dataset repository~\cite{care2026hf}. The repository provides the complete dataset, including participant metadata, annotations, and pre-computed multimodal features, together with comprehensive documentation and compressed feature archives. Programmatic access is provided through the accompanying \texttt{care-dataset} Python package, which supports metadata-based filtering, modality selection, profile-level data partitioning, and seamless integration into machine learning workflows.

\section{Code Availability}

To support reproducibility and facilitate downstream analyses on the \textbf{CARE v1.0} corpus, the full set of scripts employed to generate all multimodal descriptors and the example evaluation workflow are made available in the official Hugging Face Dataset repository alongside the dataset to ensure that results can be directly compared across studies using the same computational framework.

\bibliographystyle{naturemag}
\bibliography{main}

\section*{Funding}
\sloppy{
The work of Botelho, Trancoso, and Abad was supported by Portuguese national funds through Fundação para a Ciência e a Tecnologia, I.P. (FCT) under projects UID/50021/2025 (DOI: https://doi.org/10.54499/UID/50021/2025) and UID/PRR/50021/2025 (DOI: https://doi.org/10.54499/UID/PRR/50021/2025), and by the Portuguese Recovery and Resilience Plan and NextGenerationEU European Union funds under project C644865762-00000008 (Accelerat.AI). The work of Gimeno-Gómez and Martínez-Hinarejos was partially supported by the PROMETEO 2024 program (project LightVED, CIPROM/2023/17), forming part also of the R\&D\&I project ANNOTATE-MULTI2 (PID2024-156022OB-C32), funded by MICIU/AEI and FEDER/EU, and of the Iberian Digital Media Observatory (IBERIFIER Plus), co-funded by the EC under Call DIGITAL2023-DEPLOY-04 (Grant 101158511).
}


\section*{Acknowledgments}

We gratefully acknowledge the University of Oxford for their work in creating and sustaining the HEXI platform. We also thank Sue Ziebland and Ruth Sanders (Medical Sociology \& Health Experiences Research Group, MS\&HERG) for their availability and support. Finally, we would also like to pay tribute to the many volunteers who shared their experiences in this platform.

\section*{Author Contributions}

D. G-G., C. B., and A. A. designed the methodology.
D. G-G. and C. B collected and curated the data.
D. G-G. and C. B. wrote the initial draft version.
D. G-G. and C. B. provided the software to analyse the data.
C-D. M-H., I. T., and A. A. supervised.
C-D. M-H., I. T., and A. A. acquired the funding.
C-D. M-H., I. T., and A. A. administrated the project.
C-D. M-H., I. T., and A. A. reviewed and edited the manuscript.
All authors have read and agreed to the current version of the manuscript.

\section*{Competing Interests}

The authors declare no competing interests.

\end{document}